\documentclass{aa}
\usepackage{graphics}
\newcommand\jcd{Christensen-Dalsgaard}
\newlength{\figwidth}
\begin{document}
\thesaurus{06.01.1; 06.09.1; 06.15.1}
\title{Seismic inversions for temperature and chemical composition
profiles in the solar interior}
\titlerunning{Seismic inversions for temperature and chemical composition}
\author{H. M. Antia \inst{1} \and S. M. Chitre \inst{1,2}}
\authorrunning{Antia \and Chitre}
\offprints{H. M. Antia}
\institute{Tata Institute of Fundamental Research,
Homi Bhabha Road, Mumbai 400005, India\\
email: antia@tifrvax.tifr.res.in, chitre@tifrvax.tifr.res.in
\and
Queen Mary and Westfield College, Mile End Road, London E1 4NS, U. K.}
\date{Received }
\maketitle

\begin{abstract}

The primary inversion of the accurately measured frequencies of solar
oscillations determines the mechanical properties of the Sun, i.e.,
the sound speed and density as a function of solar radius.
In order to infer the temperature and hydrogen abundance profiles
inside the Sun, it becomes necessary to use, in addition,
the equations of thermal equilibrium
along with the input physics, such as opacities,
equation of state and nuclear reaction rates. It then becomes possible to
estimate the effects of uncertainties in opacity and nuclear reaction
rates on the inferred thermal and composition profiles.
The seismically determined temperature and
hydrogen abundance profiles turn out to be close to those in
a standard evolutionary solar model that includes the diffusion of helium
and heavy elements below the convection zone. The most significant
departure occurs just below the base of the convection zone where
the inferred hydrogen abundance profile is smoother than that in a
standard diffusion model. The region just beneath the solar convection
zone appears to be mixed, a process which could account for the observed low
lithium abundance in the solar envelope.
With a reasonable allowance for errors in opacities, the
helioseismically estimated cross-section for pp
nuclear reaction rate turns out to be $(4.15\pm0.25)\times10^{-25}$
MeV barns.

\end{abstract}

\section{Introduction}

The precisely measured frequencies of solar oscillations provide us with a
unique tool to probe the solar interior with sufficient accuracy.
These frequencies are primarily determined by the dynamical quantities
like sound speed, density or the adiabatic index of the solar material
and a primary inversion of the observed frequencies yields the sound speed
and density profiles inside the Sun (Gough \cite{dog85};
Gough \& Kosovichev \cite{dog90}; Gough \& Thompson \cite{dog91};
Dziembowski et al.~\cite{dz94}; Antia \& Basu \cite{ab94a};
Basu et al.~\cite{b96}; Gough et al.~\cite{dog96}).
On the other hand, in order to infer the temperature
and chemical composition profiles additional assumptions regarding the input
physics are required (Shibahashi \cite{shi93}; Antia \& Chitre \cite{ac95};
Shibahashi \& Takata \cite{st96}; Kosovichev \cite{kos96}).
Thus, the equations of thermal equilibrium enable us to determine
the temperature and hydrogen abundance profiles in the solar interior
provided the opacities, equation of state and nuclear energy generation
rates are prescribed. Although the primary inversions can yield the
sound speed
to an accuracy of 0.1\%, the opacities and nuclear reaction rates are
hardly known to comparable accuracy and consequently,
more systematic errors
are introduced in these secondary inversions for temperature and chemical
composition.

There are a number of approaches adopted
for secondary inversions.  Kosovichev (\cite{kos96})
has employed the equations of thermal equilibrium to express the changes
in primary variables ($\rho,\Gamma_1$) in terms of those in
secondary variables ($Y,Z$) and obtained
equations connecting the frequency differences to variations in abundance
profiles. It should be noted that modifications in $Z$
profile mainly affect the opacities
in the solar interior while the equation of state and nuclear energy
generation rates are affected to a much lesser extent. Such a
procedure is essentially equivalent to finding the $Y$
profile along with the necessary opacity modifications.
Shibahashi and Takata~(\cite{st96}, hereinafter ST96) adopt the standard
opacities and nuclear reaction rates to obtain the temperature
and chemical abundance profiles using the inverted sound speed profile.

Antia and Chitre~(\cite{ac95}, \cite{ac96}) set out
to estimate the central temperature of the
Sun. They adopted the inverted sound speed and density profiles to obtain
the temperature ($T$) and helium abundance ($Y$) profiles in the solar core,
but the main difference
was that opacity and nuclear reaction rates were not directly employed
for this purpose. Instead, the $T$ and $Y$ profiles were obtained by
minimizing the variation in opacities from the standard values.
 The main reason for allowing variations
in theoretically determined quantity like opacity rather than the
`experimentally' inferred seismic sound speed and density was that
the uncertainties
in opacities are probably larger than those in primary inversions.
Another advantage of this approach was that it enabled us to study the
effect of changes in opacity and nuclear reaction rates on the inferred
thermal profiles in a straightforward
manner.
The main thrust of the foregoing study was to estimate the central temperature
of the Sun and the neutrino fluxes. On the other hand,
Roxburgh (\cite{rox96}) and
Antia \& Chitre (\cite{ac97}) studied various possible abundance profiles
without any bounds on the opacity variations to study the implication
of helioseismic constraints on the solar neutrino problem.

In the present study we extend the earlier work of Antia and
Chitre~(\cite{ac95}) to determine
the temperature and hydrogen abundance profiles throughout the radiative
interior of the Sun
and investigate possible uncertainties that might exist
in the basic nuclear energy generation rates and opacities.
Recently, there has been a claim that the pp nuclear reaction rate should
be revised upwards by a factor of 2.9 (Ivanov et al.~\cite{iva97}) and it would
therefore be interesting to test this suggestion helioseismically
(Degl'Innocenti et al.~\cite{inn97}).
Further, in the earlier study we had restricted the composition profiles
to smooth functions represented by a low degree polynomial, which
constrained the class of admissible solutions.
This restriction has been relaxed in
the present study by using cubic spline basis functions to represent
the composition profiles.

The rest of the paper is organized as follows: the
inversion technique employed to obtain the temperature and chemical
composition profiles is described in Section 2, and the results are set out
in Section~3. Our attempts to constrain the
cross-section for the pp reaction using the inverted profiles are
outlined in Section~4, while
Section 5 summarizes the conclusions from our study.

\section{The inversion technique}

The sound speed and density profiles inside the Sun are inferred from
the observed frequencies using a Regularized Least Squares technique
(Antia \cite{a96}). The primary inversions based on the equations of hydrostatic
equilibrium along with the adiabatic oscillation equations,
however, provide only the mechanical variables like
pressure, density and sound speed. This provides us with the ratio
$T/\mu$, where $\mu$ is the mean molecular weight. In order to
determine separately $T$ and $\mu$, it becomes necessary to use the equations
of thermal equilibrium, i.e.
\begin{eqnarray}
L_r&=&-{64\pi r^2\sigma T^3\over 3\kappa\rho}{dT\over dr},\label{dtr}\\
{dL_r\over dr}&=&4\pi r^2\rho\epsilon,\label{dlr}
\end{eqnarray}
where $L_r$ is the total energy generated within a sphere of radius $r$,
$\sigma$ is the Stefan-Boltzmann constant, $\kappa$ is the Rosseland
mean opacity, $\rho$ is the density and $\epsilon$ is the nuclear energy
generation rate per unit mass. In addition, the equation of state
needs to be adopted to relate the sound speed to chemical composition and
temperature. Equation~(\ref{dtr}) is applicable when there is no convective
transport of energy. This is generally true in the region below the outer
convection zone, and we have
verified that all the models considered in this work are stable to
convection.

Since we have only three equations, namely, equations (\ref{dtr}),
(\ref{dlr}) and the equation
of state to determine the  variables $T, L_r$ and the chemical abundances,
it becomes possible to determine only one parameter specifying
the composition, e.g., the mean molecular weight.
Clearly the solution cannot be unique and therefore, in this work we
assume the heavy element abundance, $Z$ to be prescribed
and attempt to determine
the hydrogen abundance ($X$) as also the
temperature. It should be stressed that $Z$ mainly affects the opacity in the
solar interior, since the bulk of the energy generation takes place through the
pp chain. Thus, an increase in $Z$ by 20\% (which is a
reasonable estimate for possible errors in $Z$) raises the
opacity by about 8--15\%, while the sound speed changes by no more than
0.2\%, and the integrated luminosity changes by at most 0.4\%.
It is evident that the dominant effect of a change in the $Z$ profile is
on the opacities and in this work we do not make any attempt to separate
the intrinsic errors in opacity tables from those arising due to uncertainties
in the $Z$ profiles.
The opacity changes could be because of intrinsic
errors in opacity tables or due to incorrect $Z$ profiles and it
becomes difficult to separate the two effects.
One reason to keep the $Z$ profile fixed is that the value of
$Z/X$ in the convection zone is known (Grevesse \& Noels \cite{gre93}) and the
change in the interior due to diffusion is not expected to be very
large, being of the order of 10\% (Proffitt \cite{pro94}) or even less depending
on treatment of diffusion (Richard et al.~\cite{ric96}). We can thus assume
the $Z$ profile to be known to an accuracy of better than 20\%.

In order to calculate the $X$ profile we express it in terms of suitable
basis functions, e.g., B-splines, by writing
\begin{equation}
X=\sum_{i=1}^{n_s}a_i\phi_i(r),\label{xbsp}
\end{equation}
where $\phi_i(r)$ are the cubic B-spline basis functions based on
uniformly spaced knots. We use knots
with a spacing of $0.02R_\odot$, which is found to be adequate to represent
the $X$ profile to the level of accuracy expected from helioseismic
inversions. We have tried experiments by increasing the number of knots
to find that it does not have any significant effect on the solution.
For a given set of coefficients $a_i$
it is possible to calculate $X$ and then the equation of state together with
the known sound speed and density profiles determine the temperature profile.
Once the temperature, density and composition profiles
are known we can integrate 
equation~(\ref{dlr}) to calculate the luminosity, $L_r$
\begin{equation}
L_r=L_\odot-\int_r^{R_\odot} 4\pi r^2\rho\epsilon\;dr.\label{ilr}
\end{equation}
Here the upper limit of integration can be suitably truncated since
there is no significant energy generation in the outer layers. It may be
noted that this equation is integrated from the outer boundary in order
to prevent the errors in primary inversion near the center from
contaminating the results in the outer region. With this approach, the errors in
primary inversion near the center will only affect the secondary
inversion in the central region. We use the nuclear reaction rates from
Bahcall \& Pinsonneault (\cite{bp95}, hereinafter BP95) to calculate the nuclear
energy generation rate $\epsilon$.
However, the cross-section for the pp reaction which has a dominant
influence on nuclear energy generation rate
in the Sun, has decreased by about 4.5\% in the last
few years (Bahcall \cite{bah89}; BP95). We therefore, use both these
values to estimate the influence of uncertainties in the nuclear reaction
rates. Most of the calculations have been performed using the older
reaction rate for pp reaction, as that is found to give the computed
luminosity closer to the observed value, $L_\odot$.

Once the luminosity has been determined in the manner outlined above,
we can substitute
all the quantities in equation~(\ref{dtr}) to obtain
\begin{equation}
R_\kappa=-{3\kappa\rho L_r\over 64\pi r^2\sigma T^3{dT\over dr}}.\label{rk}
\end{equation}
We use the recent OPAL opacities (Iglesias and Rogers \cite{igl96}) to estimate
the ratio $R_\kappa$.
If the equations of thermal equilibrium are exactly satisfied, this quantity
would be unity everywhere, but in general that is not the case. The departures
of $R_\kappa$ from unity is a measure of the extent by which the opacity needs
to be modified to satisfy the equations of thermal equilibrium for a
composition profile prescribed by equation~(\ref{xbsp}), the coefficients
in which can then be determined by minimizing
the required opacity modifications. This outlines our
prescription for calculating the composition and temperature profiles
which may be regarded as defining a seismic solar model.
It may be noted that generally with such a procedure
the integrated luminosity will not
turn out to be equal to the observed solar luminosity.
We can adjust the nuclear energy generation rate to obtain the correct
luminosity.

In order to implement this procedure we need to measure the required
deviation of opacity.
It would be simplest to use a least squares approach, where the integral of
squared difference is minimized. Thus we minimize the quantity
\begin{equation}
F=\sum_{i=1}^n(R_\kappa(r_i)-1)^2+\alpha  ({L-L_\odot\over L_\odot})^2,\label{chisq}
\end{equation}
where $r_i$ are a set of suitable mesh points spanning the radiative
interior of the Sun, $L$ is the computed luminosity in the seismic
model and $\alpha\ge0$.
We generally use a mesh with uniform spacing of $0.005R_\odot$, which
gives approximately 145 points.
Thus we determine the $X$ profile by choosing the coefficients $a_i$
in equation~(\ref{xbsp}) to minimize $F$. 
Depending on the
value of $\alpha$, this procedure may also be able to yield the correct
solar luminosity without adjusting the nuclear reaction rates.
Thus, only the opacity deviations will be minimized for $\alpha=0$, while for
large values of $\alpha$, the integrated luminosity can also tend to
the solar luminosity at the expense of larger opacity variations.
However, in general it is found that an adjustment of the luminosity is somewhat
difficult because in the process
the resulting composition profile as well as the required
opacity modifications may become unacceptable.

In actual practice
the function is linearized about some initial guess for the $X$ profile
and least squares solution is calculated iteratively. Apart from this
simple technique we have also tried the technique of simulated annealing
(Vanderbilt \& Louie \cite{van84}; Press et al.~\cite{nr})
to obtain the nonlinear least squares fit.
Since the convergence of simulated annealing technique is very
slow, after some stage we switch to the linearized version to arrive
at the final solution.

It is not obvious that such a choice will indeed produce the correct $X$ profile
since the actual errors in opacity may be larger than the minimum estimate,
or the opacity modification may not be correctly estimated because of
errors in the primary inversion or in the nuclear energy generation
rates or in the adopted $Z$ profile.
The errors in the $X$ profile due to those
in the primary inversions can be estimated by perturbing the inverted profiles
of sound speed and density. For this purpose we repeat the primary inversions
using a perturbed set of frequencies, where randomly distributed perturbations
with variance equal to the quoted errors in frequencies are added to the
input frequencies.
However, it turns out that these errors are
fairly small as compared to those introduced by
uncertainties in opacities. For estimating the errors
arising from uncertainties in opacities we try to determine the $X$
profile for various prescribed $Z$ profiles. We use for this purpose one of the
following three basic $Z$ profiles:
(1) a homogeneous profile (denoted by HOM) without any diffusion of heavy
elements, 
(2) a profile including diffusion (Proffitt \cite{pro94}) indicated as PROF, and
(3) another profile using a different treatment of diffusion including some
turbulent mixing just below the base of the convection zone (Richard
et al.~\cite{ric96}) identified as RICH.
We scale all these
profiles to give a prescribed value of $Z$ at the solar surface,
$Z_\mathrm{surf}$ and
the inversions are performed for a very large range of values for
$Z_\mathrm{surf}$.

Inside the convection zone we can determine
the helium abundance independently (Gough \cite{dog84}; D\"appen
et al.~\cite{dap88}; Dziembowski et al.~\cite{dz91}; Kosovichev
et al.~\cite{kos92}; Antia \& Basu \cite{ab94b}; Basu \& Antia~\cite{ba95})
and the temperature can
then be determined using the inverted sound speed. However, in this
work we have restricted ourselves to the radiative interior only.
The estimated value of $X$ at the base of the convection zone can be
compared with the independently estimated values inside the convection
zone. Once the $T$ and $X$ profiles inside the Sun are known it is
straightforward to estimate the neutrino fluxes for the various
solar neutrino experiments.

\begin{figure*}
\hbox to \hsize{\resizebox{\figwidth}{!}{\includegraphics{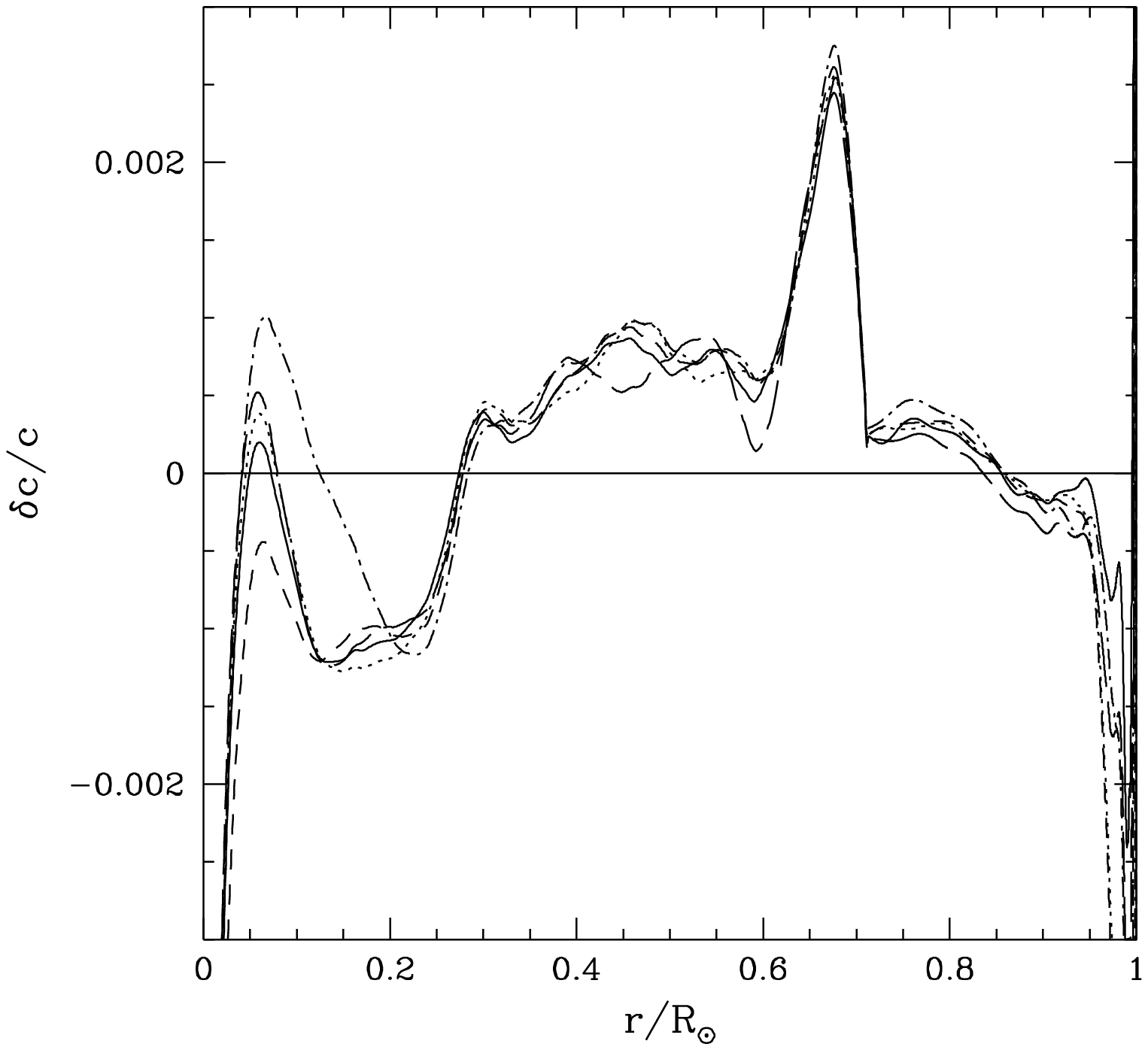}}
\hfil\resizebox{\figwidth}{!}{\includegraphics{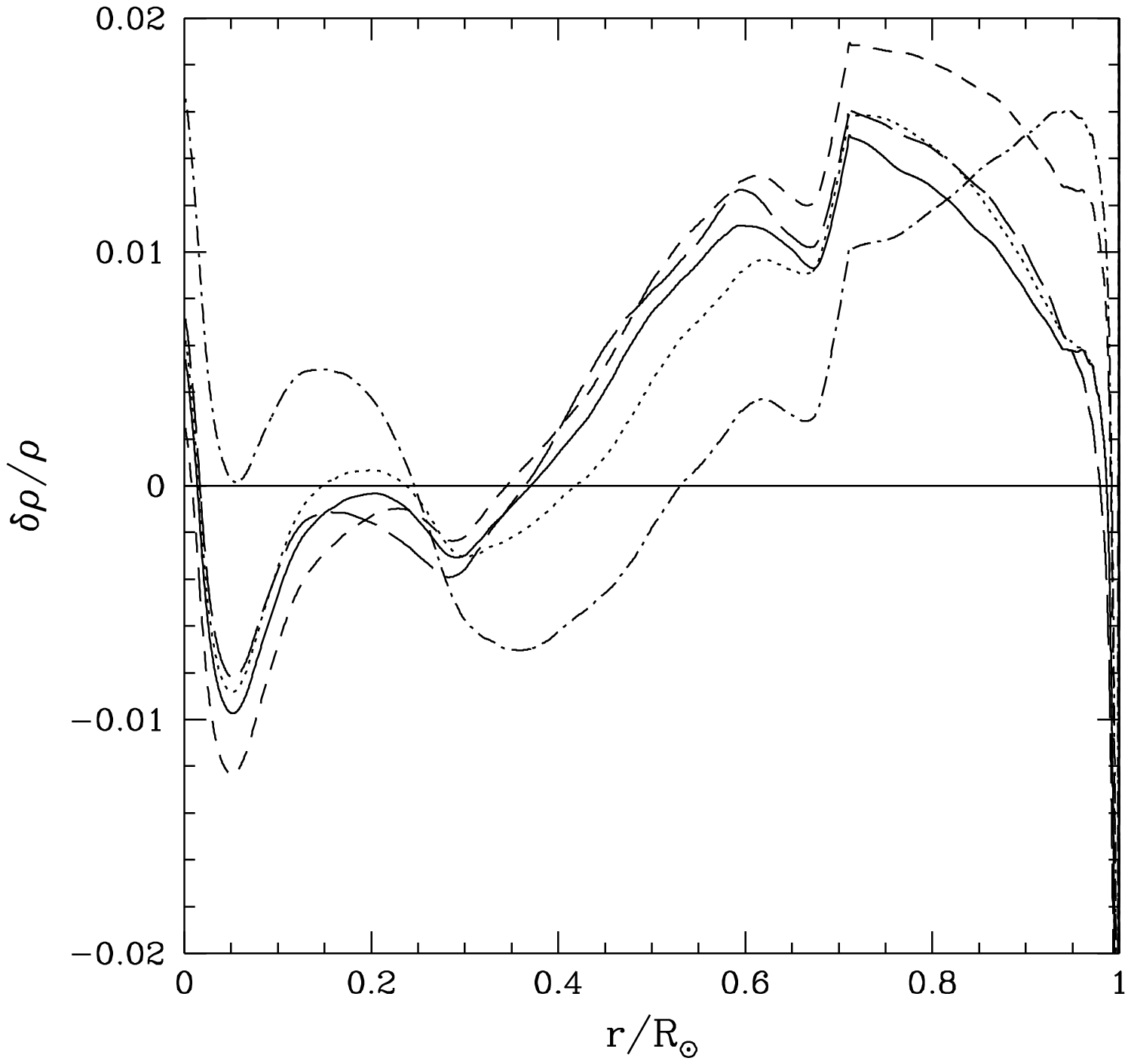}}}
\caption{Relative difference in sound speed and density
between the Sun and the Model S of \jcd\ et al.~(\cite{jcd96})
as inferred using various sets of observed frequencies.
The continuous, dotted,  short-dashed and long-dashed lines
respectively, represents the results using GONG data for the
months 4--10, month 9, months 4--7 and months 4--14. While the
dot-dashed line represents the results using BBSO data combined with
low degree frequencies from BiSON.}
\label{obsinv}
\end{figure*}

The representation~(6) takes care of both the modifications in the
opacity and nuclear energy generation rate. Thus, in the limit of
$\alpha=0$ we essentially recover the approach of ST96, for which
the required opacity modification is negligible.
When we try the least squares solution with $\alpha=0$,
the resulting opacity modifications turn out to be  generally very small and
merely represent numerical errors. We have, in fact, verified that the
two solutions are almost identical.
Most of the results in this paper have been obtained using this
prescription.
In our formulation by increasing the value of
$\alpha$ we can also find solutions which satisfy the luminosity constraint
at the expense of allowing for some opacity variations. This approach would be
similar to that of Kosovichev~(\cite{kos96}) who has estimated
$Z$ variations, which is essentially equivalent to finding the opacity
modifications.  We thus have the choice
of either modifying the nuclear energy generation rate to match the solar
luminosity ($\alpha=0$) or to modify the opacities keeping the nuclear
energy generation rate fixed ($\alpha>>1$) to achieve the same purpose.
Naturally, by using intermediate values of $\alpha$ we can obtain
solutions which require modifications of both opacities and nuclear
energy generation rates by varying amount. Clearly the resulting solution
is not unique, but all possible solutions may not be acceptable
since some of them may require unacceptably large modifications in
opacity or nuclear energy generation rates. Of course, if we can get
the observed solar luminosity for a solution with $\alpha=0$, then
effectively no modification would be required in input microphysics.
In fact, for some choices of nuclear reaction rates and opacities
we do find such solutions where no significant modifications in
microphysics are required.

It should be stressed that our technique for inferring the $T$ and $X$ profiles
in the solar interior is absolute, in the sense that no
reference model is required and the actual profiles of $T$ and $X$ are
determined directly from the sound speed and density profiles. Of course,
the density and sound speed are determined by using a differential
technique where the differences are linearized about a reference model.

\begin{figure*}
\hbox to \hsize{\resizebox{\figwidth}{!}{\includegraphics{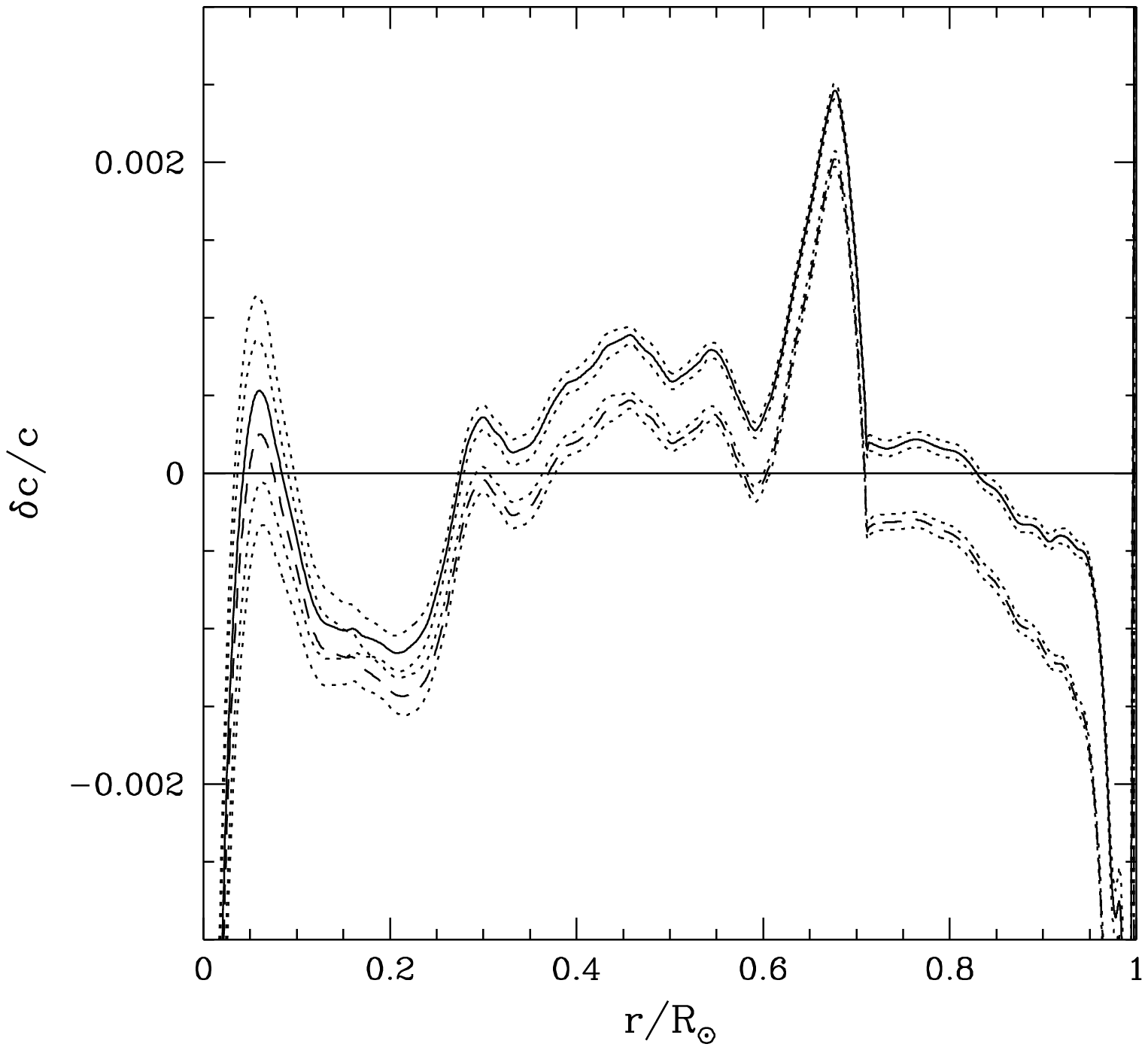}}
\hfil\resizebox{\figwidth}{!}{\includegraphics{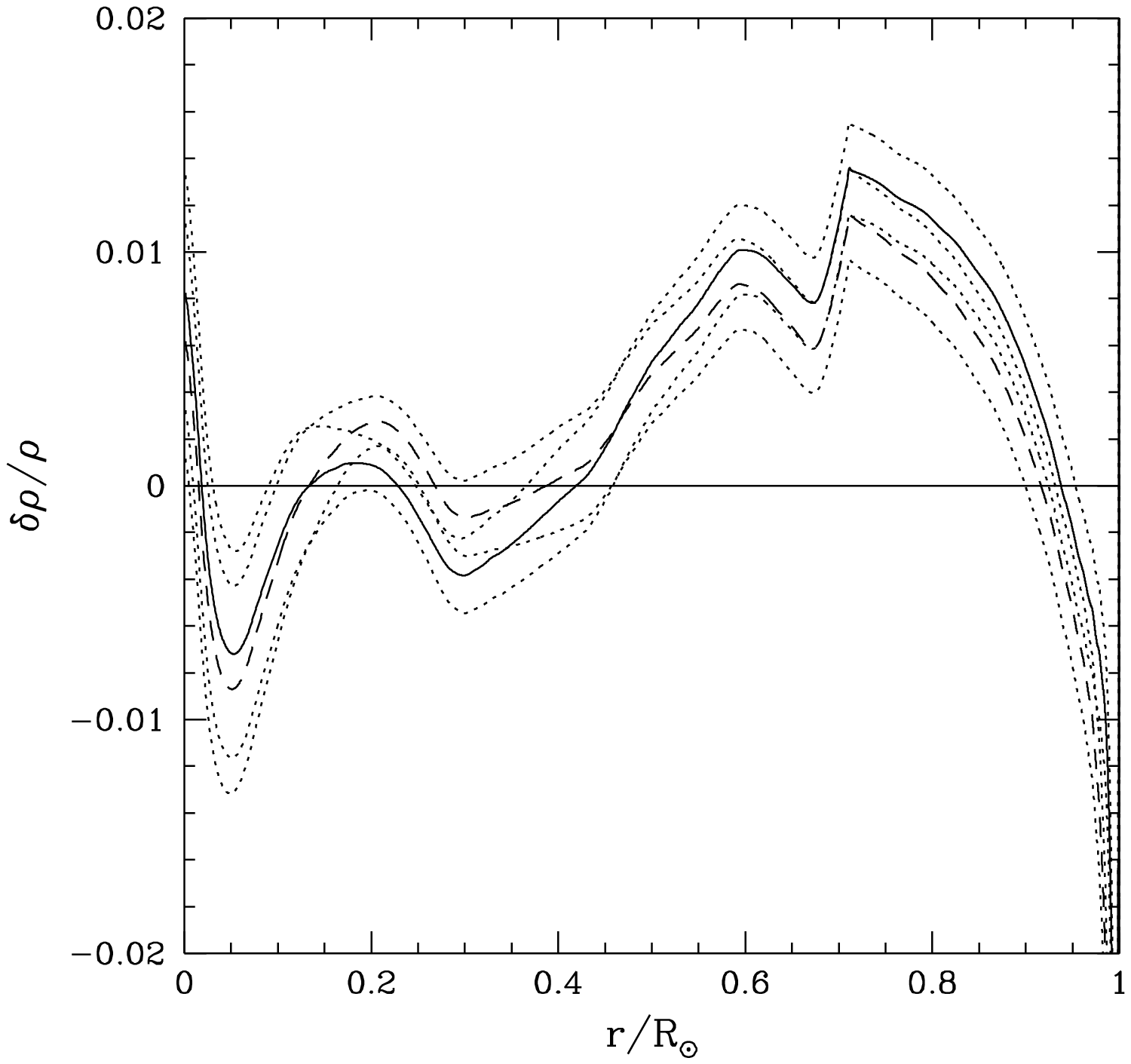}}}
\caption{Relative difference in sound speed and density
between the Sun and the Model S of \jcd\ et al.~(\cite{jcd96})
as inferred using two different estimates for solar
radius. The continuous and dashed lines represent the results obtained with
estimated radius of 695.99 and 695.78 Mm respectively. The dotted lines
show the $1\sigma$ error limits.}
\label{invrad}
\end{figure*}

\section{Inverted $T$ and $X$ profiles}

We use data sets of $p$-mode frequencies from the GONG data
(Hill et al.~\cite{hil96}) and
from the Big Bear Solar Observatory (BBSO) data (Libbrecht et al.~\cite{lib90})
along with the low degree modes from BiSON data (Elsworth et al.~\cite{els94})
to infer the sound speed and density profiles
using the Regularized Least Squares technique (Antia \cite{a96}).
The results obtained using different input frequencies
are shown in Fig.~\ref{obsinv} which displays the relative
difference between inverted profiles and those in the model S of
\jcd\ et al.~(\cite{jcd96}).
It can be seen that the results obtained using different
sets of frequencies agree with each other to within the estimated errors.
The most significant difference in the sound speed between the model and
the Sun, occurs just below the base of the convection zone and is very likely
on account of the $X$ profile in the Sun being smoother than that in the model
(Basu \& Antia \cite{ba94}; Gough et al.~\cite{dog96}; Basu \cite{b97}).
Apart from this, another noteworthy smaller hump occurs around
$r=0.2R_\odot$ which is opposite in sign to that below the convection
zone. It is likely that in this region the $X$ profile in the Sun is
steeper than that in the model. The third minor hump around $r=0.05R_\odot$
is not very significant and occurs in a region where the primary
inversions are not likely to be in any case very reliable.
Similarly, the most significant difference in the density profile
occurs inside the convection zone and is probably due to small errors
in opacity, equation of state, surface abundances and/or the depth of the
convection zone (Basu \& Antia \cite{ba97}).

Recently, it has been suggested that the standard value of solar radius
(Allen \cite{all73}) needs to be reduced (Antia \cite{a97};
Schou et al.~\cite{sch97}) and 
it would be interesting to estimate the effect of error in the solar radius on
helioseismic inversions. We have shown in
Fig.~\ref{obsinv} the results
obtained using the standard value of 695.99 Mm for the solar radius.
We have also performed inversions with a reduced radius of 695.78 Mm
and Fig.~\ref{invrad} compares the results obtained with
two different
values for solar radius using the same set of observed frequencies from
GONG months 4--10 data. It is clear that the small error in solar
radius affects the inversion results to an extent which is much larger
than the estimated errors due to those in frequencies.

\begin{figure*}
\hbox to \hsize{\resizebox{\figwidth}{!}{\includegraphics{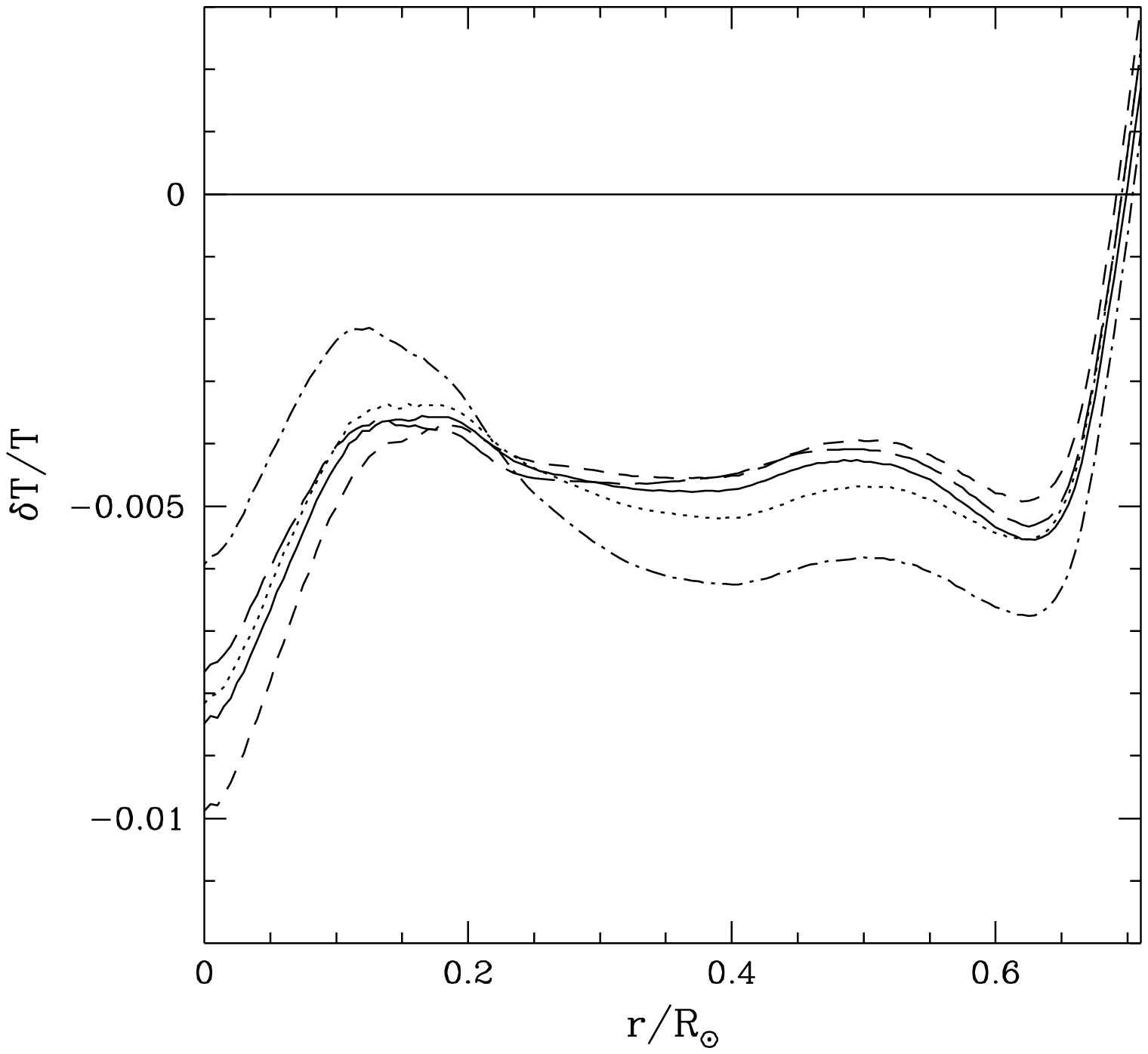}}
\hfil\resizebox{\figwidth}{!}{\includegraphics{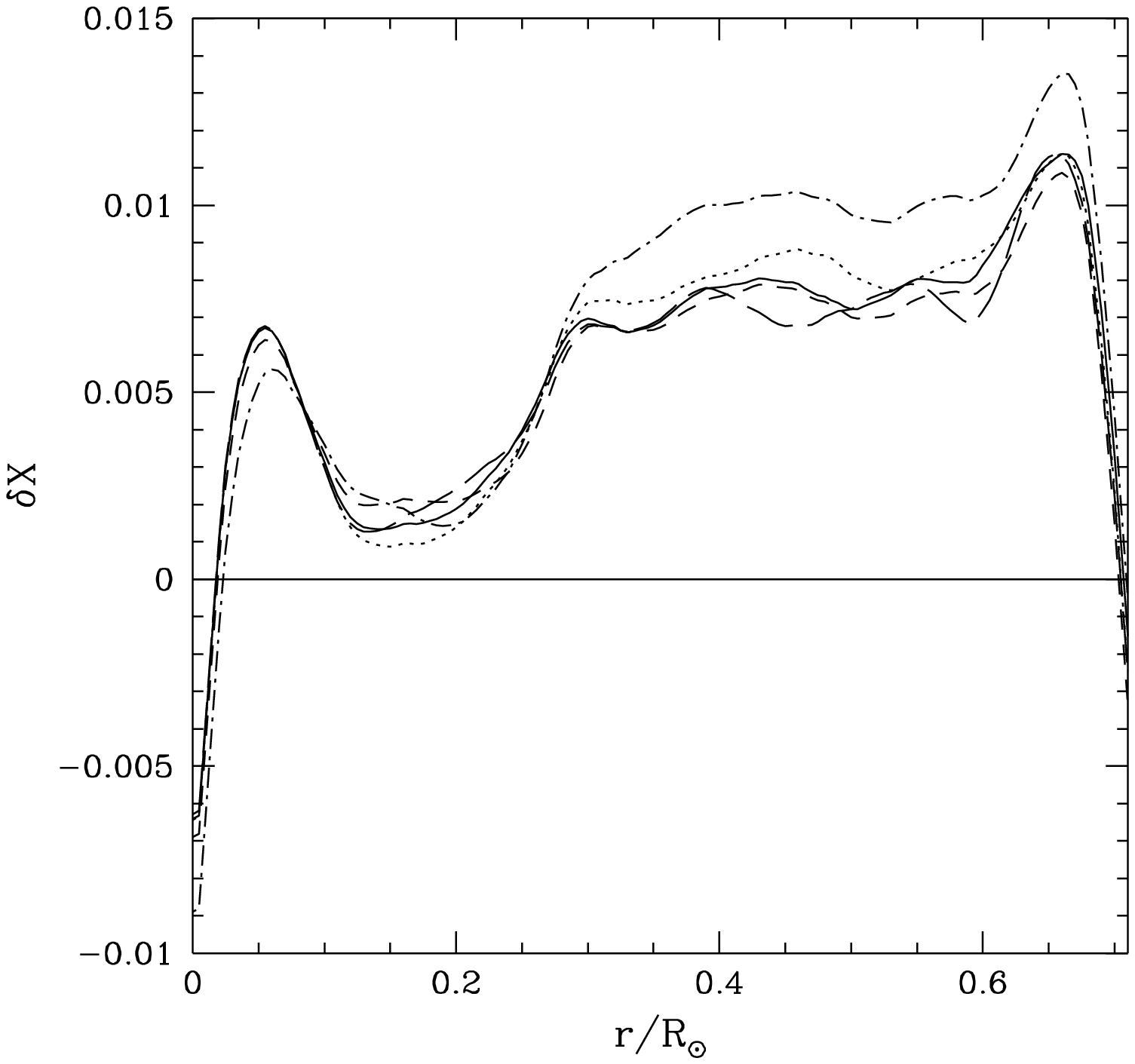}}}
\caption{Relative difference in temperature and absolute difference in
the hydrogen abundance $X$, between the Sun and the Model S of
\jcd\ et al.~(\cite{jcd96}) as inferred using different input frequencies.
The line styles are same as those in Fig.~\ref{obsinv}.}
\label{invtx}
\end{figure*}

Applying the procedure outlined in Section~2 to
the inverted profiles for sound speed and density shown in Fig.~\ref{obsinv},
we obtain the $T$ and $X$ profiles and the results are shown in
Fig.~\ref{invtx}. All these results have been obtained with $\alpha=0$ in
equation~(\ref{chisq}) and using the pp reaction rate from
Bahcall~(\cite{bah89}). A $Z$ profile including diffusion (RICH)
of heavy elements with surface value of
$Z=0.018$ was assumed for these inversions.  Once again it is clear that
the results obtained using different input frequencies
are close to one another. The computed luminosity in these seismic models
turns out to be between 0.965--0.993$L_\odot$, which is roughly consistent
with the actual solar luminosity.
The errors in secondary inversions arising
from estimated uncertainties in the input frequencies can be calculated
with a Monte-Carlo simulation. For this purpose we generate 20 sets
of  artificial
frequency data where randomly distributed errors with standard deviation
equal to the estimated errors in observed frequencies are added to every input
frequency before the primary inversion. The inverted sound speed and
density profiles are then used to obtain the $T$ and $X$ profiles.
The standard deviation in these profiles, at a fixed radius, will give an
estimate of errors in secondary inversions arising from those in input
frequencies.
It turns out that the relative errors in the inferred values of $T$
are much smaller than those in $X$, and clearly, the temperature gets
determined much more reliably than the chemical composition in this procedure.

\begin{figure*}
\hbox to \hsize{\resizebox{\figwidth}{!}{\includegraphics{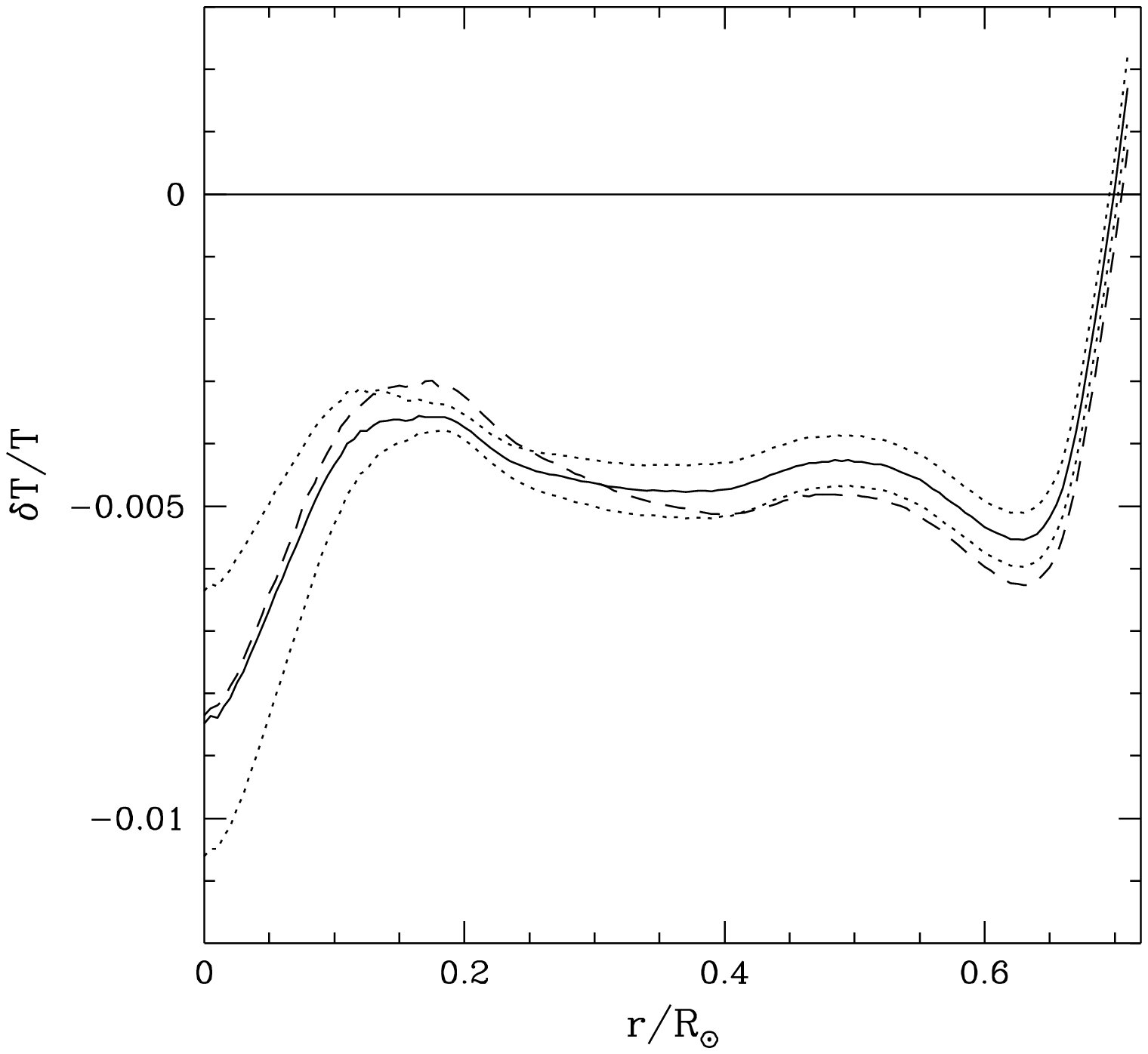}}
\hfil\resizebox{\figwidth}{!}{\includegraphics{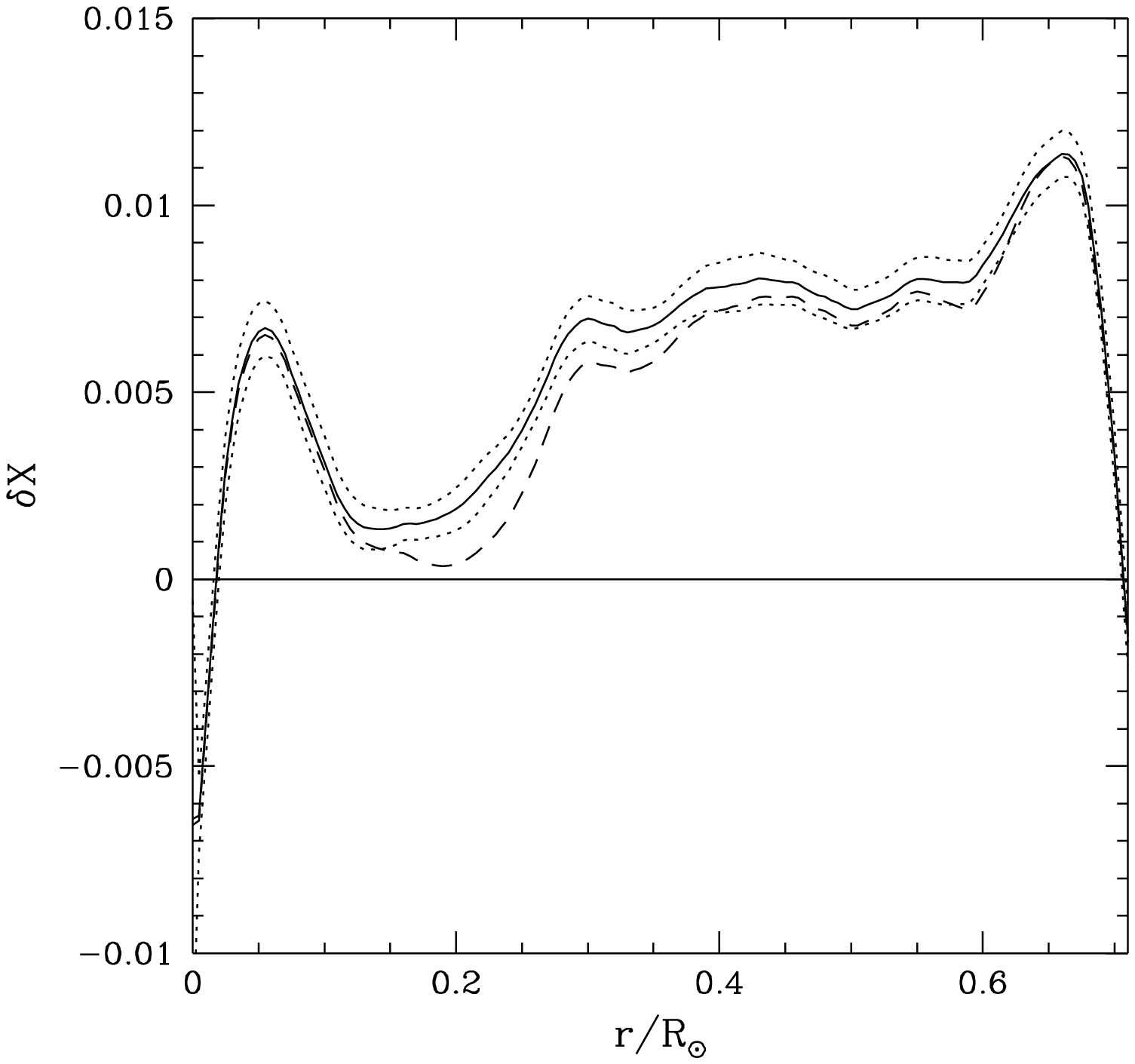}}}
\caption{Relative difference in temperature and absolute difference in
the hydrogen abundance $X$, between the Sun and the Model S of
\jcd\ et al.~(\cite{jcd96}) as inferred using different estimates for
solar radius.
The continuous and dashed lines represent the results obtained with
estimated radius of 695.99 and 695.78 Mm respectively. The dotted lines
show the $1\sigma$ error limits on the results with estimated radius of
695.99 Mm.}
\label{txrad}
\end{figure*}

We can also estimate the influence of error in adopted solar radius on
the secondary inversions by using sound speed and density profiles obtained
with different values of the radius and these results are shown in
Fig.~\ref{txrad}. These errors can be seen to be comparable to those
due to uncertainties in frequencies.
The properties of these seismic models are summarized in Table~\ref{tab1},
which also gives the estimated errors in each quantity due to those in
the GONG months 4-10 data. In this table $T_c$ is the central temperature,
$\phi(^{37}\mathrm{Cl})$ the neutrino flux in the Chlorine detector,
$\phi(^{71}\mathrm{Ga})$ the neutrino flux in the Gallium detector,
while $\phi(^8\mathrm{B})$ is the flux of $^8$B neutrinos.
It can be seen that a reduction in radius by 210 km increases the
computed luminosity by $0.004L_\odot$ and the neutrino fluxes are also 
correspondingly
enhanced by similar amounts. The last line in the table gives the
results for a static solar model designated as INV, which is
constructed using the inverted $X$ profile. This model will be discussed
later in this section.
In the following discussion all the results have been obtained using the
GONG months 4--10 data with an estimated radius of 695.78 Mm
(Antia \cite{a97}).

\begin{table*}
\caption[]{Properties of seismic models}
\label{tab1}
\begin{tabular}[]{lcccccccc}
\hline
\noalign{\smallskip}
$R_\odot$& $Z$-profile& $Z_\mathrm{sur}$& $X_\mathrm{surf}$& $T_c$& $L/L_\odot$&
$\phi(^{37}\mathrm{Cl})$& $\phi(^{71}\mathrm{Ga})$& $\phi(^8\mathrm{B})$ \\
(Mm)& & & & ($10^6$ K) & & (SNU)& (SNU)& ($10^6$ cm$^{-2}\mathrm{s}^{-1}$)\\
\noalign{\smallskip}
\hline
\noalign{\smallskip}
695.99& RICH & 0.018 & 0.7349 & 15.54 & 0.970 & 6.57 & 124.0 & 4.69 \\
695.78& RICH & 0.018 & 0.7348 & 15.54 & 0.974 & 6.62 & 124.6 & 4.72 \\
695.78& PROF & 0.015 & 0.7612 & 15.40 & 0.964 & 5.64 & 118.8 & 3.92 \\
695.78& PROF & 0.018 & 0.7347 & 15.59 & 0.977 & 7.04 & 127.0 & 5.07 \\
695.78& PROF & 0.020 & 0.7195 & 15.72 & 0.985 & 8.09 & 132.7 & 5.94 \\
695.78& HOM & 0.015 & 0.7613 & 15.30 & 0.959 & 5.04 & 115.3 & 3.44 \\
695.78& HOM & 0.018 & 0.7348 & 15.48 & 0.971 & 6.21 & 122.4 & 4.39 \\
695.78& HOM & 0.020 & 0.7196 & 15.59 & 0.979 & 7.08 & 127.4 & 5.10 \\
\noalign{\smallskip}
\multicolumn{3}{l}{Error estimates}& 0.0008 & \phantom{1}0.03 & 0.010 & 0.30 & \phantom{10}2.3 & 0.24 \\
\noalign{\smallskip}
695.78& INV & 0.018 & 0.7351 & 15.57 & 1.000 & 7.23 & 127.3 & 5.28 \\
\noalign{\smallskip}
\hline
\end{tabular}
\end{table*}

Apart from this there could be some errors due to uncertainties in the equation
of state resulting from the use of inappropriate values of $Z$.
These errors could be estimated using the approximate expression
for the correction to the sound speed, namely,
\begin{equation}
{1\over c^2}\left(\partial c^2\over\partial Z\right)_{T,\rho,X}=
-{1\over 3+5X-Z}. \label{dcz}
\end{equation}
This error is found to be very small as compared to other uncertainties.
It may be noted that if we take the partial derivative at constant
$Y$ instead of constant $X$, the right hand side of equation~(\ref{dcz}) will
be 6 times larger and the
corresponding error will also be larger. In this work we therefore attempt
to determine the $X$ profile rather than $Y$ profile from secondary
inversion. With this choice the errors due to uncertainties in $Z$
affecting the equation of state are negligible. Of course, the
uncertainties in the equation of state itself will also affect the results,
which can be estimated by repeating the inversion procedure with a different
equation of state. Thus, while most of the results were obtained using the
OPAL equation of state (Rogers et al.~\cite{rog96}), we have also done
some inversions using the MHD equation of state (D\"appen et al.~\cite{mhd1};
Hummer \& Mihalas \cite{mhd2}; Mihalas et al.~\cite{mhd3}). It turns out that
the difference in these results is not significant and the computed luminosity
decreases by $0.002L_\odot$ when the MHD equation of state is used instead
of OPAL.  It would seem that
the secondary inversion results are not particularly sensitive to
reasonable uncertainties in the equation of state.

In principle, for any given $Z$ profile it should be
possible to find $T$ and $X$ profiles which do not require any opacity
modification as has been demonstrated by ST96.  In fact, a choice of
the parameter  $\alpha=0$ in equation (\ref{chisq}) produces a
profile which requires very little opacity variation and is similar
to what is obtained if we were to use the procedure adopted by ST96.
However, in this case it is not possible to ensure that the computed
luminosity will necessarily match the observed solar luminosity
$L_\odot$. This follows directly from the equations
of thermal equilibrium (Eqs.~(1,2)). Once the sound speed, density
and $Z$ profiles are known it is possible to integrate these equations
to calculate the $T$ and $L_r$ profiles, which will depend only on
the central temperature $T_c$. Clearly, by adjusting $T_c$ it is not
possible to get both the correct luminosity and $Y$ at the base of the
convection zone. In fact, the solution is so sensitive to the choice of
$T_c$ that a change in $T_c$ by mere 1000 K, results in the value of
$Y$ in the convection zone to increase from 0.25 to a value
greater than 1! Thus, it is not possible to alter the luminosity even by
$0.001L_\odot$ by adjusting $T_c$ in the allowed range.
Hence, the difference between the computed and the observed
solar luminosity will give an estimate of uncertainties in primary inversions
or the input $Z$ profile, or the basic microphysics, such as the equation
of state, opacities and the nuclear reaction rates. It may be difficult
to separate out the contributions from each of these sources. We will
try to examine this question in some detail in the following section.

\begin{figure*}
\hbox to \hsize{\resizebox{\figwidth}{!}{\includegraphics{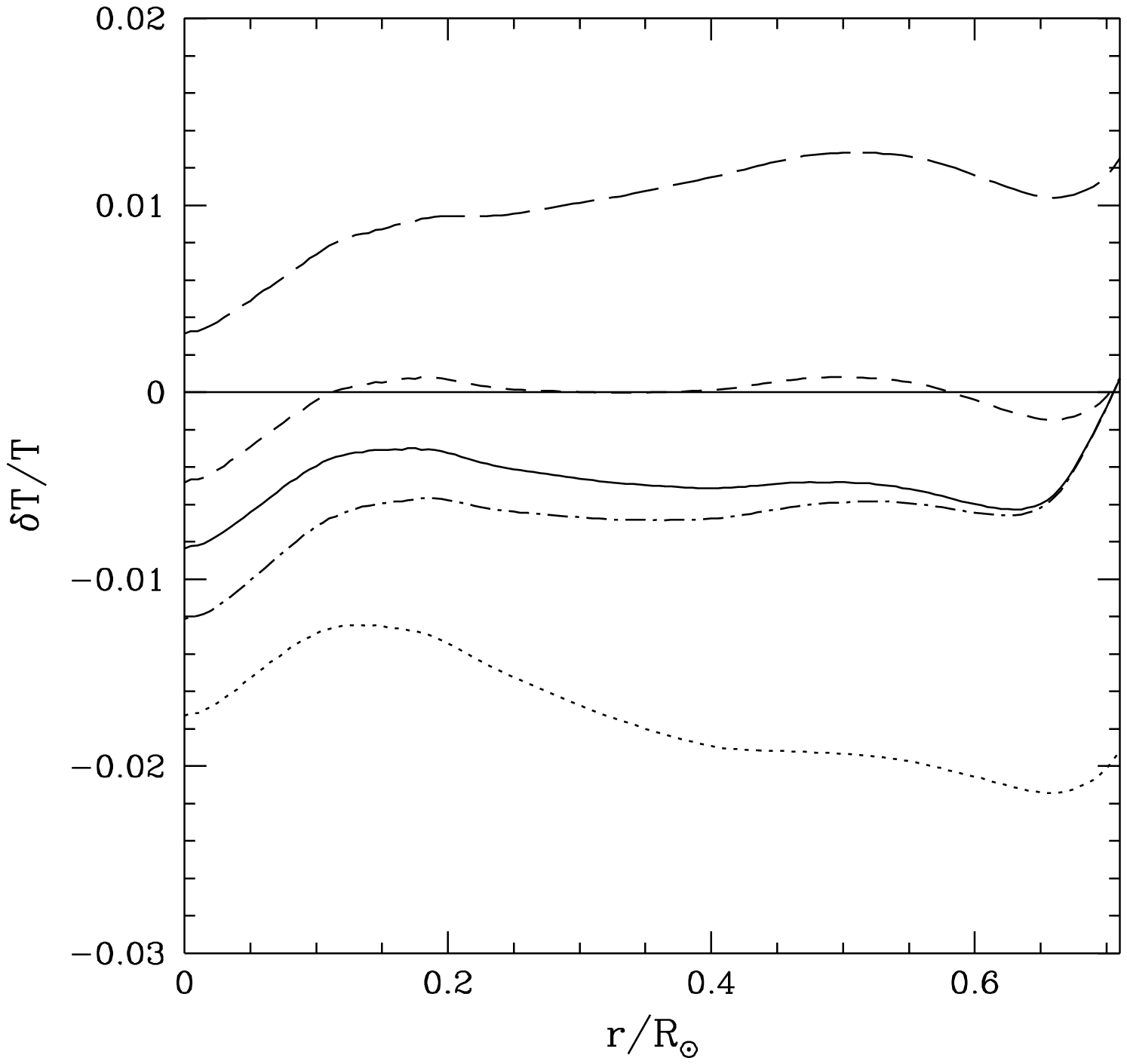}}
\hfil\resizebox{\figwidth}{!}{\includegraphics{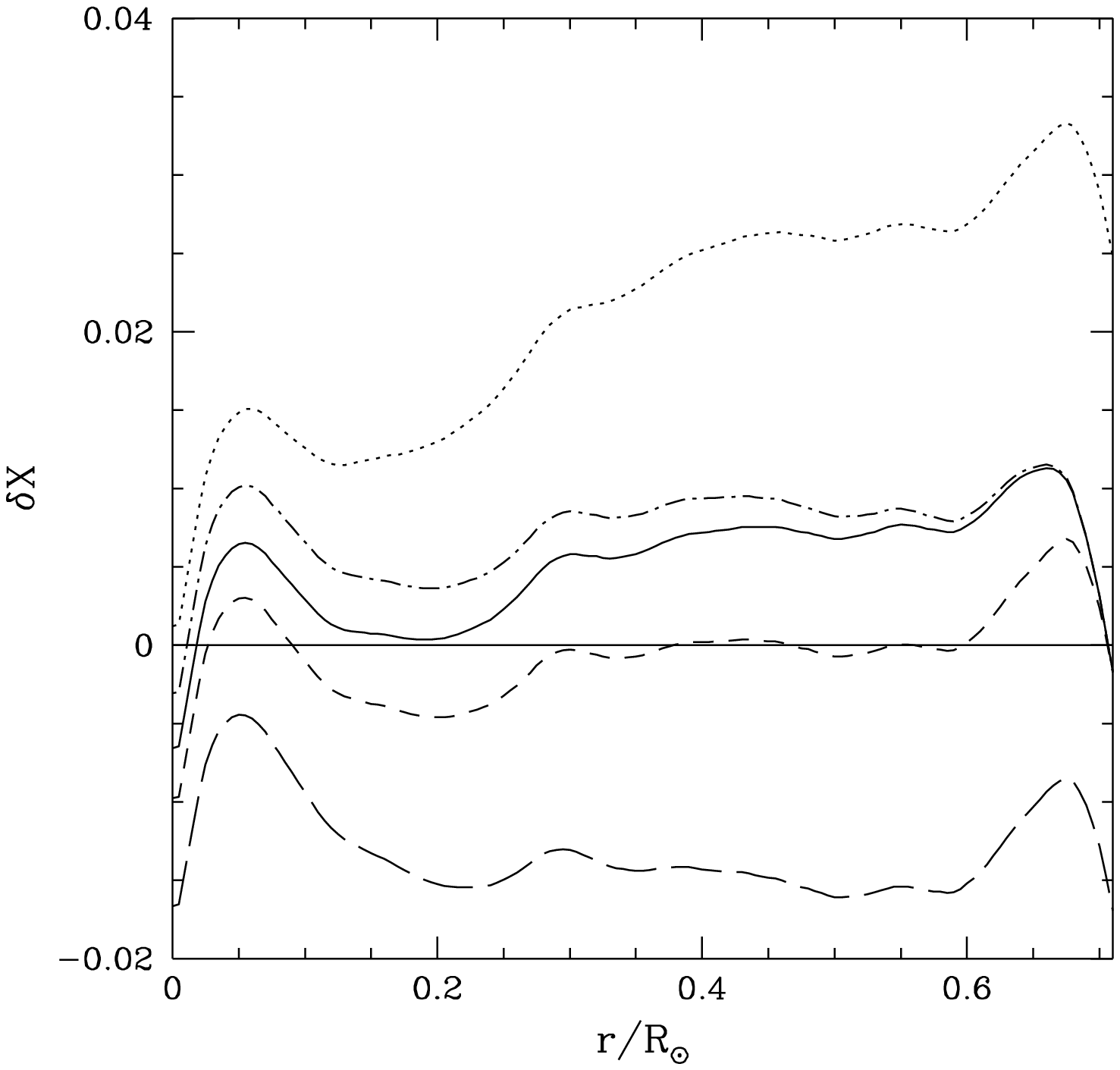}}}
\caption{Relative difference in temperature and absolute difference in
the hydrogen abundance $X$, between the Sun and the Model S of
\jcd\ et al.~(\cite{jcd96}) as inferred using different $Z$ profiles.
The continuous line represents the results using the profile RICH,
with $Z_\mathrm{surf}=0.018$. The dotted ($Z_\mathrm{surf}=0.015$),
short-dashed ($Z_\mathrm{surf}=0.018$) and long-dashed
($Z_\mathrm{surf}=0.02$) lines shows the results obtained using the
profile PROF, while the dot-dashed line displays the results using
a homogeneous $Z$ profile, with $Z=0.018$.}
\label{invz}
\end{figure*}

Of course, the inferred $T$ and $X$ profiles will depend on the assumed
$Z$ profile. In order to estimate this effect, we try a few, different
input $Z$ profiles and the results are shown in Fig.~\ref{invz}.
It is clear that  uncertainties in these profiles arising from those in
$Z$ are much larger than those due to other effects considered earlier.
Further, these uncertainties decrease with $r$, because the effect of
$Z$ on opacities decreases as temperature increases because of increasing
degree of ionization of heavy elements. There is a large uncertainty
in regions immediately below the convection zone; however, in this region
the value of $Z$ is more reliably known from the measured value in the
convection zone. Interestingly, the inverted temperature profile obtained
using the $Z$ profile PROF with $Z_\mathrm{surf}=0.018$ comes close
to that in Model~S of \jcd\ et al.~(\cite{jcd96}).
This is very likely because, with
similar $Z$ profiles, the opacities and hence the temperature gradient
should be similar between two models. The variation in the sound speed is
mainly reflected in the difference in $X$ profile.
The discordance between various profiles in Fig.~\ref{invz} could
give an estimate of errors expected from reasonable errors in $Z$ profile.

\begin{figure}
\resizebox{\figwidth}{!}{\includegraphics{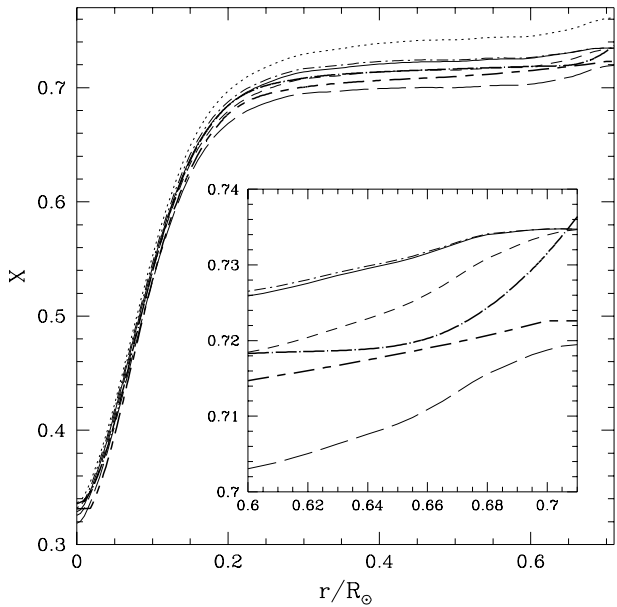}}
\caption{The hydrogen abundance profile
as inferred using different $Z$ profiles.
The various line styles have the same representation
as those in Fig.~\ref{invz}. In addition the $X$ profiles in the
Model~S of \jcd\ et al.~(\cite{jcd96}) is shown by heavy dot long-dashed line,
while that in the model~5 of Richard et al.~(\cite{ric96}) is shown by
the heavy short-dashed long-dashed line.}
\label{xabs}
\end{figure}

The absolute $X$ profiles as inferred using different $Z$ profiles are
shown in Fig.~\ref{xabs}, which also displays the profiles in some standard
solar models with different treatment of diffusion.
It is evident from this figure that the $X$ profile just
below the convection zone is much smoother than that in a standard
solar model with conventional treatment of diffusion (\jcd\ et al.~\cite{jcd96})
suggesting that some turbulent mixing probably takes place in this region
(Richard et al.~\cite{ric96}).
The $X$ profile in Model~5 of Richard et al.~(\cite{ric96})
which includes turbulent diffusion is closer to the inverted profiles,
though it appears to be shifted below the inverted profiles using
$Z_\mathrm{surf}=0.018$, probably because it has larger
$Z_\mathrm{surf}=0.019$ (and correspondingly higher $Z/X$). The shape of
the $X$-profile near the base of the convection zone is essentially
independent of $Z_\mathrm{surf}$, but depends on the actual profile used.
Thus, if a flat $Z$ profile like HOM or RICH is used the resulting $X$ profile
is also flat until about $r=0.68R_\odot$ indicating that this region
is essentially mixed.
However, if a $Z$ profile with steep gradient near the base of
the convection zone is used, then the resulting $X$ profile also shows some
weak gradient in that region. But in order to get a gradient as steep
as that in the $X$ profile of Model~S, one requires a $Z$ gradient
which is about 5 times that in the profile PROF. Hence, if the
$X$ and $Z$ profiles from similar treatment of diffusion in a solar model
are used, the resulting profiles will not be consistent with helioseismic data
unless some process like turbulent diffusion is employed to reduce
the gradients to zero
at the base of the convection zone (Basu \& Antia \cite{ba95};
Basu \cite{b97}). We
prefer to use inverted profiles with zero gradient in $X$ or $Z$
at the base of the convection zone for better accordance. 

Notice, around
$r=0.25R_\odot$ the $X$ profile in the Sun is steeper than that in the solar
model. This can be seen more clearly from Fig.~\ref{invtx} which shows the
difference
in $X$ profile between the Sun and a solar model. The steep positive gradient
around $r=0.25R_\odot$ indicates that the $X$ profile in the Sun is steeper
than that in the model.

\begin{figure}
\resizebox{\figwidth}{!}{\includegraphics{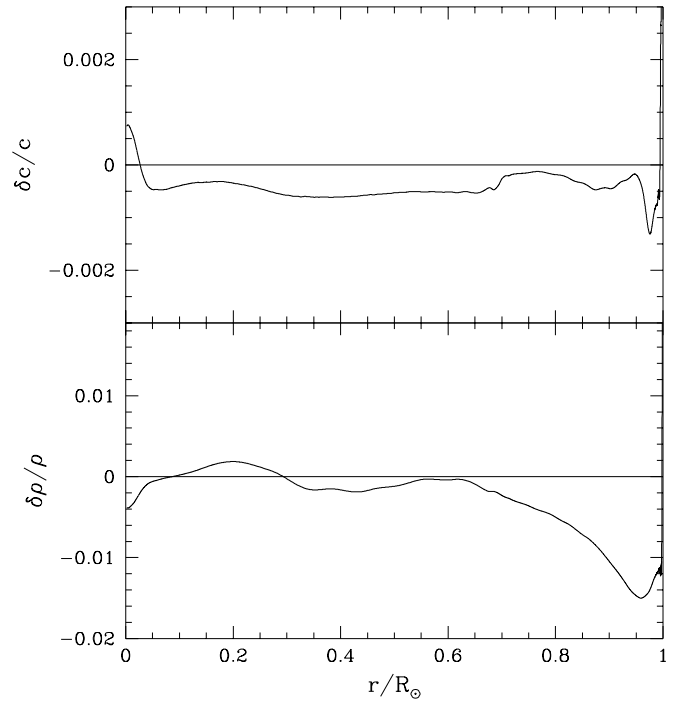}}
\caption{Relative difference in sound speed and density
 between the Sun and Model INV.
 The model INV has been constructed with the inverted
profile of hydrogen abundance shown by the continuous line in Fig.~\ref{xabs}.}
\label{invm}
\end{figure}

In order to verify the seismically inferred composition profile,
we have constructed a static solar model (Model INV)
using the inverted profile for $X$ as shown by the continuous line
in Fig.~\ref{xabs} and the  model so computed is compared with results from
primary inversions.
This model also uses the OPAL equation of state (Rogers et al.~\cite{rog96})
and opacities (Iglesias \& Rogers \cite{igl96}) and nuclear reaction
rates from BP95, except for the pp reaction for which the cross-section
estimated in the Section~4 is used.
Since the surface hydrogen
abundance $X_\mathrm{surf}$ and the mixing length parameter in these models are
adjusted to get the correct radius and luminosity, the $X$ profile
has to be scaled by multiplying it by a constant factor
and as such the resulting model does not have the
same abundance profile as that inferred from inversion. 
Fig.~\ref{invm} shows the relative difference in sound speed and density
between these models and the Sun, while the properties of this model are
also summarized in Table~\ref{tab1}.
It is clear that the hump below the
convection zone has more or less vanished.
The discrepancy in sound speed within the convection zone is likely
to be due to uncertainties in the equation of state or the error in
estimated radius, or may arise from errors in
inversion due to influence of surface layers (Antia \cite{a95}).
It can be seen that for this model the sound speed
and density are very close to those in the Sun, more or less within the
estimated errors in primary inversions.
Some of the remaining differences could be attributed to
errors in opacities, equation of state or nuclear reaction rates,
which have not been adjusted while constructing these models or on account
of errors in primary inversions. It may be noted that the neutrino
fluxes in this model as well as other seismic models listed in Table~1
are significantly lower than those in the standard solar model of BP95
with diffusion
of helium and heavy elements, and marginally higher than those in the
solar model of Turck-Chi\'eze \& Lopes~(\cite{tc93}), which does not
incorporate any diffusion of elements.

\section{Helioseismic estimate for the pp reaction rate}

The nuclear energy generation in the solar interior is mainly
controlled by the pp reaction rate.
The theoretically estimated cross-section for this reaction varies from
$3.89\times10^{-25}$ MeV barns (BP95) to $4.21\times10^{-25}$ MeV barns
(Turck-Chi\'eze \& Lopes \cite{tc93}).
For convenience we denote the usually accepted value (Bahcall \cite{bah89}) by
$S_0=4.07\times10^{-25}$ MeV barns, and express the cross-section
in terms of $S_0$.
Recently, there has been a claim that the pp nuclear reaction rate
should be revised upwards by a factor of 2.9 (Ivanov et al.~\cite{iva97}).
Although this claim has been contested (Bahcall \& Kamionkowski \cite{bah97})
on the nuclear physics grounds, it would be nice to have an independent check
from helioseismic data (Degl'Inn\-ocenti et al.~\cite{inn97}).
The inverted profiles
for $T$ and $X$ can be used to compute the total luminosity generated
in the seismic models provided the nuclear reaction rates are assumed to
be known. With the requirement that the Sun is in thermal equilibrium,
the computed
luminosity should agree with the observed value and that would impose
some constraint on the nuclear reaction rates.
However, as we have seen in the previous section,
the inverted profiles depend on the assumed profile for heavy element
abundance. We therefore, investigate the effect of an assumed $Z$ profile
on integrated luminosity to constrain the nuclear reaction rate.

Apart from the $Z$ profile, there could also be uncertainties in
the theoretically calculated values of opacities. In order to
obtain constraints which are independent of errors in opacity we can
consider  $X$ profiles, with coefficients in equation~(\ref{xbsp})
chosen arbitrarily. These arbitrary profiles may not satisfy the equations of
thermal equilibrium with any reasonable estimate for opacities.
However, using the inverted sound speed and assuming the $X$ and $Z$ profiles,
it is possible to calculate the temperature profile inside the Sun.
Once these thermal and composition profiles are known,
the luminosity in corresponding
seismic models can be computed. In order to estimate an upper limit
on the pp nuclear reaction rate we can try to construct a profile which
generates the minimum energy for the given sound speed and density
profiles. Since the
sound speed essentially constrains the value of $T/\mu$, where $\mu$ is
the mean molecular weight of the solar material, it seems in order
to cut down the energy generation one should reduce $T$ as well as
$\mu$ to keep the
ratio constant. It is clear that the minimum value of $\mu$ is achieved
when $X=1$ and $Z=0$, i.e. when there is no helium or heavy elements
present in the central region. From more
detailed calculation of energy generation rate, we have verified that this
is indeed true, although strictly speaking, since the temperature is
not high enough for helium burning reactions, the minimum energy
generation occurs when $X=0$, when there is no fuel to burn!
But even a value of $X=0.005$ gives much higher energy
generation rate as compared to $X=1-Z$ in the core,
because the temperature has to
be increased when $X$ decreases to keep the sound speed constant.
Further, if the temperature is required to decrease monotonically
with radial distance, then even such profiles can be ruled out.
Thus, leaving aside this unlikely possibility, the minimum
energy generation occurs when $X=1$ and $Z=0$, when there is no
helium in the core ($Y=0$).

For the case of a profile with $X=1$ and $Z=0$
we can easily demonstrate that the computed luminosity in the resulting
seismic model is about $0.617L_\odot$ when the usual nuclear reaction
rates are adopted. Now if we increase the pp nuclear reaction rate for
obtaining the correct solar luminosity with this profile, it turns out that
the cross-section needs to be increased to about $1.62S_0$.
It is clear that if the cross-section is increased beyond this value
it is not possible to find any $X$ profile (apart from the one where
hydrogen is almost totally exhausted throughout the solar core),
which will simultaneously
yield the correct sound speed and luminosity in seismic models.
The exact limiting value of the cross-section will depend on
the inverted sound speed  and density
profiles, but as we have seen in the previous section these uncertainties
are very small. We can therefore, conclude that
any value higher than $1.65S_0$ is inadmissible even if arbitrary
errors in opacities are allowed and the Sun is assumed to generate the
observed luminosity. An increase in the pp nuclear
reaction rate by a factor of 2.9 (Ivanov et al.~\cite{iva97}) is certainly
ruled out by the
helioseismic data. In fact, in actual practice even the profile
with $Y=0$ considered in obtaining this limit is unacceptable since
one would expect significant amount of helium to be present in the solar core.
If we consider a profile with $Y=0.2$, which is still lower than
the expected helium abundance, the limiting cross-section for the pp reaction
drops to $1.27S_0$. It is therefore evident that, any significant increase in
the pp cross-section is demonstably inconsistent with helioseismic constraints.
We would like to add that there is no straightforward way to set a lower bound
on this cross-section from such an analysis, as by increasing the helium
abundance suitably it is possible to reproduce the solar luminosity even
when this cross-section is significantly reduced, although such profiles
may require inadmissibly large opacity modifications.

In the foregoing discussion we have allowed for arbitrary errors in standard
opacity tables. Even though such an analysis helps in illustrating that the
helioseismic data are able to put severe constraints on nuclear reaction rates,
the resulting bounds on cross-section are highly conservative and are
unlikely to be achieved in realistic situations. It would be possible
to obtain more meaningful bounds if one allows only reasonable errors
in opacities. There are two problems with this approach; first,
it is difficult to define what is a reasonable error in opacity and
second, the error in opacities may have arbitrary variation with
temperature and density, thus making it difficult to consider all possible
variations even within the assumed limits. One possibility is to use
the procedure outlined in the Section~2 with a suitably large value of
$\alpha$ in equation~(\ref{chisq}), to obtain the $X$ profile which generates
the correct luminosity and requires some minimum opacity variation for
any specified nuclear reaction rates. We consider this approach later, but
before that we adopt a simpler procedure by taking different $Z$ profiles
with a large range of $Z_\mathrm{surf}$ to see how the
computed luminosity varies with $Z$. In this process, the opacity changes are
accounted through changes in $Z$ profiles.

\begin{figure}
\resizebox{\figwidth}{!}{\includegraphics{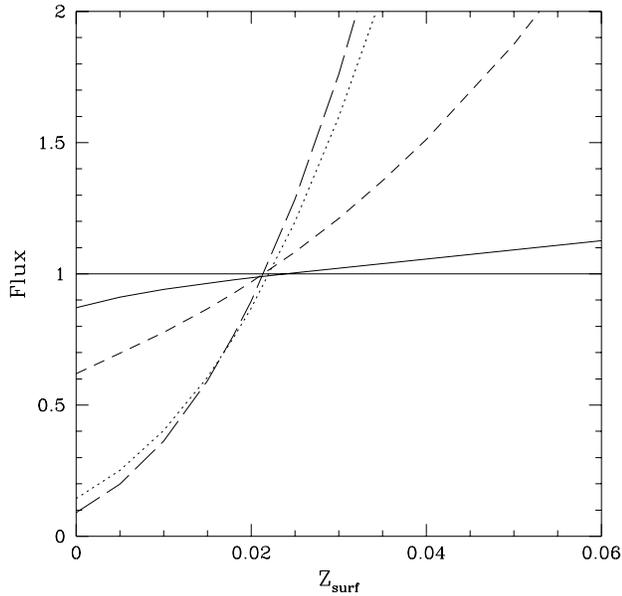}}
\caption{The integrated luminosity and neutrino fluxes for seismic models
as a function of $Z_\mathrm{surf}$. All these quantities are scaled
with respect to the values in the standard solar model of BP95.
The continuous line shows the computed luminosity, dotted line shows
the neutrino flux in the chlorine experiment, short-dashed line shows
the neutrino flux in the Gallium experiment while the long-dashed
line shows the $^8$B neutrino flux.}
\label{lum}
\end{figure}

Using the $Z$ profile with diffusion (Proffitt \cite{pro94}) scaled to different
values of $Z_\mathrm{surf}$ we can calculate the $X$ profiles
following the procedure outlined in Section~2, with $\alpha=0$ in
equation~(\ref{chisq}). The total luminosity and neutrino fluxes in the
resulting seismic models are shown in Fig.~\ref{lum}. It is clear that
the integrated luminosity goes up with $Z_\mathrm{surf}$ as a result of
increase in opacities, but not very significantly -- a variation of
$Z_\mathrm{surf}$ from 0 to 0.06, results in an increase in the
luminosity from $0.87L_\odot$ to $1.13L_\odot$. The range
of $Z$ values covered by these models is in all probability much more
than the expected uncertainties in the $Z$ profile. With the allowance of
a factor of
two variation in $Z_\mathrm{surf}$, one gets an error of about
5\% in computed luminosity. An uncertainty by a factor of two is
probably the most that is expected in $Z$, and hence we have only considered
profiles where $Z$ is scaled uniformly by the same factor. A slightly
smaller change in $Z$ in selected regions may also give rise to similar
change in the resultant luminosity.
From the results presented later where we try to
adjust the $Z$ profile to match the luminosity, it turns out that the
required maximum change in $Z$ is not much smaller than what is indicated by
this simple analysis. For the purpose of this work we therefore estimate
a reasonable error of 5\% in the luminosity arising from possible
uncertainties in the heavy element abundance and/or opacities. Since
this is much larger than the estimated uncertainties from other sources
we assume a total uncertainty of 5\% in computed luminosity. There are
uncertainties in other nuclear reaction rates which will also affect the
computed luminosity, but again if these are within limits given by
BP95, the error in luminosity from these is only about 1--2\%.
Thus the integrated luminosity is consistent with the observed value
within these uncertainties for a reasonable $Z$ profile.
It may be noted that all these results are obtained using the pp reaction
cross-section to be $S_0$. If the recent value adopted by BP95
($0.9558S_0$)
was used, the computed luminosity would be about 4\% lower, while for the
normal value of $Z$ the computed luminosity would be significantly lower than 
the observed value. This leads us to surmise that the cross-section for the
pp-reaction rate needs to be increased to its earlier value given by
Bahcall (\cite{bah89}). Similar conclusions were also reached earlier by 
Antia \& Chitre (\cite{ac95}).

\begin{figure}
\resizebox{\figwidth}{!}{\includegraphics{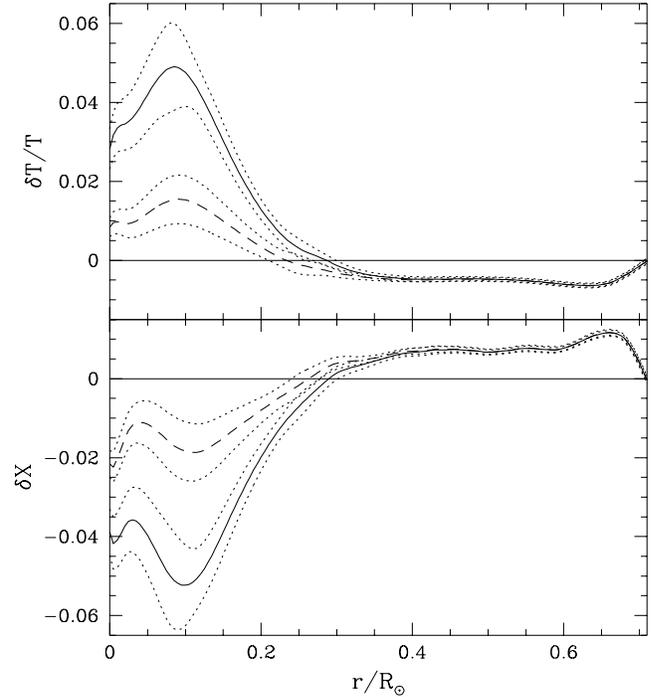}}
\caption{The relative difference in temperature and absolute difference
in sound speed between the Sun and Model S of \jcd\ et al.~(\cite{jcd96}).
These results have been obtained with the nuclear reaction rates from BP95
with the $Z$ profile also adjusted to yield the correct luminosity.
The continuous line shows the results obtained with pp reaction rate from
BP95, while the dashed line shows the results obtained when pp reaction
rate from Bahcall~(1989) is used. The dotted lines denote the $1\sigma$
error limits due to those in the input frequencies from GONG months 4--10
data.}
\label{invlum}
\end{figure}

In order to obtain a better estimate for the cross-section of pp reaction,
we try to compute the luminosity using different values for the
cross-section of the pp reaction, with the normal value of
$Z_\mathrm{surf}=0.018$. From these results we can identify the range of
cross-section values which yield the computed luminosity within 5\% of
the observed value.  This can be treated as the helioseismic estimate
for the cross-section of pp reaction, which turns out to be
$(4.15\pm0.25)\times10^{-25}$ MeV barns, where the quoted errors
correspond to an uncertainty of 5\% in the computed luminosity.
This range is consistent with
the value adopted by Bahcall~(\cite{bah89}), but slightly larger than the
more recent value adopted by BP95.

\begin{figure}
\resizebox{\figwidth}{!}{\includegraphics{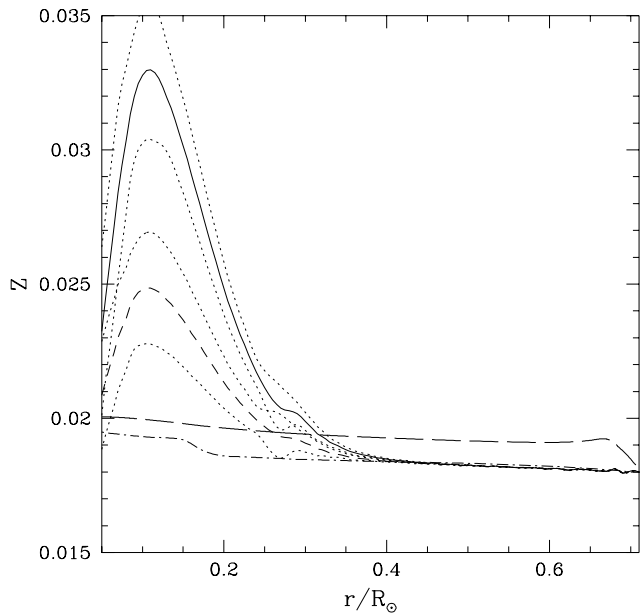}}
\caption{The $Z$-profile needed to obtain correct computed luminosity
using the nuclear reaction rates adopted from BP95. The continuous line
shows the required $Z$ profile when pp reaction rate from BP95 is used,
while the short dashed line shows the results obtained when pp reaction
rate form Bahcall~(1989) is used. The dotted lines denote the $1\sigma$
error limits due to those in the input frequencies from GONG months 4--10.
The long dashed  and the dot dashed lines respectively,
show the profiles PROF and RICH scaled to $Z_\mathrm{surf}=0.018$.}
\label{zinv}
\end{figure}

All the inversion results presented so far were obtained using
$\alpha=0$ in equation~(\ref{chisq}), which yield profiles that require
no opacity modifications, but the computed luminosity may not match
the observed value. It is possible to adjust the opacity or
equivalently the heavy element abundance to obtain the correct observed
luminosity by choosing a suitably large value of $\alpha$ in
equation~(\ref{chisq}).
However, such profiles may not be unique as only one parameter namely, the
luminosity is fitted by adjusting the $Z$-profile in radiative interior.
Nevertheless, we can obtain a possible solution which yields a seismic
model with correct luminosity. We use the nuclear reaction rates as
adopted by BP95 (including that of pp reaction) for this study.
Then the computed luminosity turns out to be
about $0.94L_\odot$ for the case of $\alpha=0$ (i.e., no opacity
modifications), but if we choose a large value of $\alpha$, say $2500$,
the computed luminosity comes out to be close
to the observed value. The resulting value of $R_\kappa$ at
each point can then be converted to equivalent variation in $Z$, and this gives
the $Z$ profile required to obtain the correct total solar luminosity.
For this purpose we use the technique of simulated annealing to minimize
the function $F$ defined by equation~(\ref{chisq}). Once the iteration
is close to convergence we linearize about that solution and determine
the actual $X$ profile.
The resulting $T$ and $X$ profiles are shown in Fig.~\ref{invlum},
while Fig.~\ref{zinv} shows the required $Z$ profile. These figures
also show the estimated errors arising from those in frequencies.
If the cross-section for the pp nuclear reaction is taken to be $S_0$
then the required increase in integrated luminosity is much smaller and
the resulting profiles are also shown in Figs.~\ref{invlum} and \ref{zinv}.
It is clear
from these figures that with the BP95 reaction rates
the profiles need to be modified significantly
to generate the extra luminosity and the resulting $Z$ profile in
the central regions does
not look particularly realistic, as there is an unrealistic hump in the
core.
Thus, it appears that $Z$ needs to be increased
by about a factor of two to obtain the correct luminosity when
the nuclear reaction rates as adopted by BP95 are used.
If the $Z$ profile is uniformly scaled by a constant factor (cf.,
Fig.~\ref{lum}) of two, there will be a similar change in the luminosity.
Further, the estimated errors are also larger than those in
Fig.~\ref{invtx}, because the profiles are more sensitive to
errors in primary inversions when the luminosity constraint is applied.
When the pp nuclear reaction rate is adopted from Bahcall~(1989)
the resulting $Z$ profile still has the same shape as before, but the
height of the hump is much less and it is within $2\sigma$ of the
usual $Z$ profiles including diffusion.

The sensitivity of inverted profiles to nuclear reaction rates
is probably due to helioseismic constraints which are being applied
in this work. Since the density and sound speed are known from primary
inversion, in order to maintain the
observed solar luminosity with the reduced nuclear reaction rates,
the temperature will need to be increased.
The sound speed constraint from the primary inversions fixes the ratio
$T/\mu$, where $\mu$
is the mean molecular weight. The mean molecular weight $\mu$ will
also have to be increased, implying a decrease in the hydrogen abundance $X$.
This will work against enhancing the energy production, and
as a result, temperature will have to be increased substantially
to keep up with the required nuclear energy production for maintaining
the observed luminosity.
If we allow for the departures from the inverted sound speed and density
profiles then it may be
possible to obtain the correct luminosity without much modification
in $Z$, but the resulting seismic model will not have the inferred sound
speed in the core.

\section{Discussion and Conclusions}

In this work we have shown that the use of sound speed and density profiles
obtained from primary inversions enables us to infer the temperature
and hydrogen abundance profiles, provided the heavy element abundance
profile as well as the microphysics like the equation of state, opacities
and nuclear energy generation rates are known. The profiles so determined
would define a seismic model, although the integrated luminosity in these
seismic models may not necessarily match the observed value. The difference
may arise due to uncertainties in primary inversions, and/or the assumed
$Z$ profile, and/or the microphysics. While it is possible to adjust the $Z$
profile to yield the observed luminosity, it is not clear if that is
the correct solution, since the discrepancy may arise due to other reasons.
We have attempted to estimate the extent to which various uncertainties
can influence the luminosity to find that the effect of 
equation of state or primary inversions on computed luminosity to be
fairly small. The dominant uncertainty arises from the nuclear
reaction rates and opacities (or equivalently the $Z$ profile).
It is difficult to separate out the influence of these two factors, but
if we assume a reasonable error in one of these the other effect can
be quantified.

It turns out that if we use the nuclear reaction rates adopted by BP95,
except for the pp reaction for which the older reaction rate from
Bahcall~(\cite{bah89}) is used, then the integrated luminosity with the normal
$Z$ profile is close to the observed value. It is thus tempting to
conclude that these
nuclear reaction rates, together with the current opacity tables and
a $Z$ profile including diffusion are consistent with helioseismic data.
Similarly, from a detailed study
of the base of the convection zone it appears that uncertainties in the
current opacities at the base of the convection zone
as well as the estimated $Z/X$ values (Grevesse
\& Noels \cite{gre93}) are fairly small (Basu \& Antia \cite{ba97}).
One expects opacities to
be more reliably determined in the solar core where temperatures
are upwards of several million degrees. It is therefore reasonable to
assume that there are no significant uncertainties in current OPAL
opacity tables in the solar core. Of course,
there could be some error on account of an inappropriate $Z$ diffusion profile,
but that is not expected to be too large.

It is remarkable that, different values of cross-section for the pp-reaction
have been adopted by various workers (Bahcall \cite{bah89}; BP95;
Turck-Chi\'eze \& Lopes \cite{tc93}; Dar \& Shaviv \cite{dar96}) and recently
Ivanov et al.~(\cite{iva97}) have even suggested an increase
in the pp reaction rate by a factor of 2.9.
Since there is no experimental measurement of this cross-section, it would
be interesting to indulge in an exercise to estimate this cross-section
helioseismically.
From our results in the previous section it is clear that an increase in
this reaction rate by a factor of 2.9 is essentially ruled out, even when
arbitrary variations in opacities are allowed.
The only $X$ profiles which may yield the computed luminosity as low
as the observed value with such nuclear reaction rates are those where
hydrogen is almost completely exhausted ($X<0.002$) in most of the core.
In fact, even an increase by a factor of 1.65 in the pp reaction rate
is inconsistent with helioseismic data, with no
restriction on opacity. If the helium abundance is constrained
to a minimum of 0.2, then this limiting factor is decreased to 1.3.
Thus we can firmly conclude that even a 30\% increase in the
cross-section for pp reaction is inconsistent with helioseismic data.
However, these bounds are too
conservative since unrestricted errors in opacity are permitted.

Should we make the assumption, on the other hand
that opacities are known to reasonable accuracy and that there is
an uncertainty of up to a factor of
two in $Z_\mathrm{surf}$, then this translates into an uncertainty of 5\%
in computed luminosity and the estimated value of the cross-section
for the pp reaction turns out to be $(4.15\pm0.25)\times10^{-25}$ MeV barns.
This value is consistent with the estimate of $4.07\times10^{-25}$ MeV barns
(Bahcall \cite{bah89}; Dar \& Shaviv \cite{dar96}) or
$4.21\times10^{-25}$ MeV barns (Turck-Chi\'eze \& Lopes \cite{tc93}),
but slightly larger than the value of $3.89\times10^{-25}$
MeV barns adopted by BP95. It thus appears that the estimate of the
pp reaction cross-section adopted by BP95,
needs to be increased by a few percent. With the adoption
of the recent estimate of this cross-section, the
$Z$ profile will need to be modified by about a factor of two to obtain the
correct computed luminosity. We cannot, of course, strictly rule out such
$Z$ profiles, but they appear unlikely to be realized in practice.

The reliability of the inverted seismic profiles from the observed frequencies
can be demonstrated by constructing a model (INV) with the inverted
$X$ profile. This model is found to be close to observations in
many respects including the sound speed and density profiles through the
solar interior. Thus,
the base of the convection zone in model INV is found to be at a radial
distance of
$0.7131R_\odot$, which is consistent with the helioseismically estimated
value (\jcd\ et al.~\cite{jcd91}; Basu and Antia \cite{ba97}). Similarly,
the helium abundance in the envelope of this model is found to be
0.2469, which is also close to the helioseismically estimated value
(Basu and Antia \cite{ba95}). This appears to indicate that
the inverted $X$ profile used in this model is close to that in the Sun.
Note that this model has been constructed using the standard
OPAL opacities and equation of state.
It should be recognized that this model satisfying the seismic constraints
is probably not unique, as it may be possible to construct different solar
models satisfying the helioseismic constraints by modifying the opacities
or nuclear reaction rates or the $Z$ profiles suitably.

The inverted $T$ and $X$ profiles are found to be close to those in the
Model S of \jcd\ et al.~(\cite{jcd96}).
It should be stressed that our technique is
absolute in nature, and hence the fact that the resultant
inverted profiles are close to those
of a standard solar model cannot be a coincidence, since the model profiles
have not been used anywhere during the secondary inversion. Our
results, therefore, appear to demonstrate that the temperature and hydrogen
abundance profiles in the Sun are close to those in a standard solar model.
The major noticeable difference arises just below the base of the convection
zone, where the inverted $X$ profile is smoother than that in the 
standard model. The $X$ profile is in fact, sensibly flat in the
region $r>0.68R_\odot$.  This is probably owing to some process involving
turbulent diffusion just below the base of the convection zone,
which is not accounted for in the usual treatment of diffusion
(Richard et al.~\cite{ric96}).  Such a
mixing could smoothen the composition gradient and also explain
the anomalously low lithium abundance in the solar photosphere.
It should be stressed that the estimated uncertainties due to
errors in the $Z$ profile are fairly large and consequently,
significance of the flatness of the profile may not be obvious.
However, a mere increase or
decrease in the opacity by a constant factor will not change the nature
of the profile as similar results can be obtained for different values of
$Z_\mathrm{surf}$.
Only if there is a sharp gradient in modified opacity over this
narrow region (or equivalently a sharp gradient in the $Z$ profile)
it will be possible to obtain composition profiles which
are not flat just below the convection zone. If the gradient in $Z$
profile were to be increased by a factor of five over that in
Proffitt~(\cite{pro94}),
it would be possible to get an $X$ profile with gradient similar to that
in Model~S at the base of the convection zone. Thus, composition
profiles obtained
using similar treatment of diffusion for both helium and heavy elements
are not consistent with inverted profiles unless the gradient vanishes
as in the case of turbulent mixing (Richard et al.~\cite{ric96}).
These results are consistent with conclusions drawn
from the oscillatory signal in the frequencies
(Basu \& Antia \cite{ba94}; Basu \cite{b97}), which also supports
the presence of turbulent mixing in this region.
Similar evidence is also suggested by the inversion
of sound speed (Gough et al.~\cite{dog96}). All this seems to indicate that
the region just below the convection zone is probably mixed
(Richard et al.~\cite{ric96}) by some process.

In contrast, in the central region around $r=0.25R_\odot$ the
composition profile in the Sun
appears to be steeper than that in the solar model, perhaps suggesting
that mixing is unlikely to have occurred in this region of solar interior.
From Fig.~\ref{invtx} it can be seen that $\delta X$ has a negative gradient
in the inner core around $r=0.1R_\odot$, which would imply that the
$X$ profile in the Sun is smoother than that in the model. This difference
has presumably been considered as a hint of mixing in the core
(Gough et al.~\cite{dog96}).
However, considering the fact that the $X$ gradient is very steep in this
region, the difference is extremely small and mixing if any, could only
have taken place in the very early history of solar evolution or the
mixing process is extremely slow. A more likely
cause of this difference is the errors in nuclear reaction rates.
It is also possible that this difference could arise from uncertainties
in the primary inversion in the core. 

Using the inverted $T$ and $X$ profiles it is possible to estimate
the neutrino fluxes. From the results in Table~1, it appears that
these neutrino fluxes are significantly lower than those in the standard
solar model of BP95, with diffusion of helium and heavy elements.
Some of the difference could be due to somewhat lower cross-section for
pp reaction used by BP95. A part of the difference will also arise
from the diffusion of heavy elements. As argued earlier there are
good reasons to believe that the region immediately below the convection
zone is mixed and hence the heavy element abundance will not increase as
steeply as in the model of BP95. A reduction in $Z$ value inside the core
will reduce the opacities and hence the temperature and the corresponding
neutrino fluxes. However, the computed neutrino fluxes in seismic models
are significantly larger than the observed values. In fact, it has been
found (Antia \& Chitre \cite{ac97}) that even if arbitrary variations in
opacities are allowed it is not possible to reduce the neutrino
fluxes in any two solar neutrino experiments simultaneously to the observed
values. Thus, it appears that the solution of solar neutrino problem should
be sought in terms of neutrino properties, though the seismic models
can be used to constrain these solutions. Since the neutrino fluxes in
the standard solar model of BP95 are somewhat different from those in
the seismic models, the constraints on the particle physics solution
(e.g., Hata \& Langacker \cite{hat97}) could change when seismic models
are used.

We have demonstrated that our inversion technique produces reasonably
well the thermal and composition profiles in the Sun's interior, with
the knowledge of the sound speed and density inferred from
the accurately observed frequencies, based on the mechanical and thermal
equilibrium constraints governing the solar structure. These
seismically determined temperature and hydrogen abundance profiles in the
Sun turn out to be close to those obtained with a standard solar
model. The small departures could be due to a variety of processes
arising from diffusion and uncertainties in nuclear reaction rates,
equation of state, heavy element abundances or even the presence of a
magnetic field. It is remarkable that the neutrino fluxes in the
framework of the seismic model come out to be close to those predicted
by the standard solar model, assuming that the opacities are not very
different from the currently accepted OPAL values. There is thus a strong
hint of the particle physics solution of the solar neutrino puzzle!

\begin{acknowledgements}
This work utilizes data obtained by the Global Oscillation
Network Group (GONG) project, managed by the National Solar Observatory, a
Division of the National Optical Astronomy Observatories, which is 
operated by AURA, Inc. under a cooperative agreement with the 
National Science Foundation.
The data were acquired by instruments operated by the Big
       Bear Solar Observatory, High Altitude Obseratory,
       Learmonth Solar Observatory, Udaipur Solar Observatory,
       Instituto de Astroflsico de Canarias, and Cerro Tololo
       Interamerican Observatory.
We should like to thank I. W. Roxburgh
for valuable discussions.
\end{acknowledgements}

\end{document}